\documentclass{article}
\usepackage{spconf,amsmath,amssymb,graphicx}
\usepackage[table]{xcolor}
\usepackage{booktabs,array,multirow,placeins}
\usepackage{hyperref}

\definecolor{metricgray}{HTML}{ECECEC}
\newcommand{\method}{EmoGavel}

\newcommand{\Lcal}{\mathcal{L}}

\title{EmoGavel: Grounding-Aware Reinforcement Learning with Verifiable Rewards for Multimodal Emotion Preference Learning}

\name{\shortstack{Zilong Huang\textsuperscript{1}, 
Kong Aik Lee\textsuperscript{1}, Zezhong Jin\textsuperscript{1}, Man-Wai Mak\textsuperscript{1}, Tatsuya Kawahara\textsuperscript{2}}}
\address{\shortstack{\textsuperscript{1}The Hong Kong Polytechnic University, Hong Kong SAR, China\\
 \textsuperscript{2}Kyoto University, Japan}}

\begin{document}
\ninept
\maketitle

\begin{abstract}
Multimodal emotion preference learning uses pairwise comparisons to
identify which open-ended emotion descriptions align more closely with
human judgments. However, existing methods primarily focus on the final
preference decision and provide little explicit supervision for
localizing and correcting erroneous content. To address this limitation, we propose \textbf{EmoGavel}, a \textbf{G}rounding-\textbf{A}ware reinforcement learning framework with \textbf{V}erifiable rewards for multimodal \textbf{Emo}tion preference \textbf{L}earning. We first introduce grounded supervised fine-tuning (Grounded SFT), which exploits controlled-error construction information to jointly learn preference prediction, error localization, repair, and missing-evidence recovery. We then optimize these verifiable outcomes with grounding-aware group relative policy optimization (GA-GRPO), combining preference-gated grounding rewards with dynamic zero-variance rescue to prioritize preference correctness and recover learning signals from all-wrong groups. Experiments on EmoPrefer and controlled-error evaluation sets show that EmoGavel improves both natural preference prediction and controlled-error discrimination while maintaining strong error-grounding performance. Our code and model checkpoints will be released once the paper is accepted.

\end{abstract}

\begin{keywords}
multimodal emotion understanding, preference learning, reinforcement learning with verifiable rewards
\end{keywords}

\section{Introduction}
\label{sec:intro}

Descriptive Multimodal Emotion Recognition (DMER) expresses human emotional states using free-form natural language, enabling fine-grained descriptions beyond conventional categorical emotion recognition~\cite{lian2023explainable,huang2026emoeus,cheng2024emotionllama,lian2026mer,huang2026eiiscl,chaubey2026avere,mm-nodeformer}. However, evaluating such open-ended descriptions is challenging because open-ended emotion descriptions allow multiple valid expressions and are difficult to evaluate against a single reference~\cite{lian2026emoprefer,gao2026beyond,wang2026emotionthinker}. EmoPrefer formulates the task as pairwise preference prediction, where a multimodal large language model (MLLM) judge~\cite{chen2024mllmjudge,tan2025judgebench,wang2024unfair} selects the description that better reflects the target person's emotional state~\cite{lian2026emoprefer}. Such preference supervision can identify more accurate descriptions, but fail to reveal which content in the alternative description is emotionally inconsistent with the emotional state or how the description should be corrected.

Previous work typically trains an MLLM judge from pairwise preference labels using supervised fine-tuning (SFT) or direct preference optimization (DPO)~\cite{christiano2017deep,ouyang2022instructgpt,rafailov2023direct_duplicate2}. Conventional pairwise supervision is often sparse, typically pairing each preferred description with only one negative example. Error-Augmented Preference Optimization (EAPO) enriches this supervision by constructing multiple controlled negative descriptions from each preferred description~\cite{huang2026eapo}. However, these controlled pairs are still optimized primarily for binary preference prediction, without explicitly teaching the model where an error occurs or how it should be corrected. Consequently, a model may learn to select the preferred description without developing a corresponding ability to identify and correct the specific erroneous content~\cite{xing2026emotionhallucer,wang2025audio}.

We observe that controlled-error construction contains richer supervision information than the resulting preference label alone. Since each controlled negative is produced by modifying a preferred description, the construction process naturally preserves the correspondence between the original and corrupted contents. To exploit this supervision, we propose EmoGavel, a grounding-aware reinforcement learning framework with verifiable rewards for multimodal emotion preference learning. EmoGavel consists of two stages: Grounded SFT first establishes error-grounding capability from construction-derived supervision, followed by our \emph{Grounding-Aware GRPO (GA-GRPO)} for direct optimization of verifiable preference and grounding outcomes. GA-GRPO incorporates preference-gated grounding rewards and dynamic zero-variance rescue to preserve preference correctness and recover learning signals from all-wrong output groups.

Our main contributions are as follows. First, we extend multimodal emotion preference learning beyond binary selection by introducing explicit supervision for error localization, repair, and missing-evidence recovery. Second, we propose EmoGavel, a two-stage framework that integrates Grounded SFT with Grounding-Aware GRPO for joint preference and error-grounding optimization. Third, we introduce preference-gated grounding rewards and dynamic zero-variance rescue to improve preference robustness under verifiable reinforcement learning.

\begin{figure*}[t]
\centering
\includegraphics[width=\textwidth]{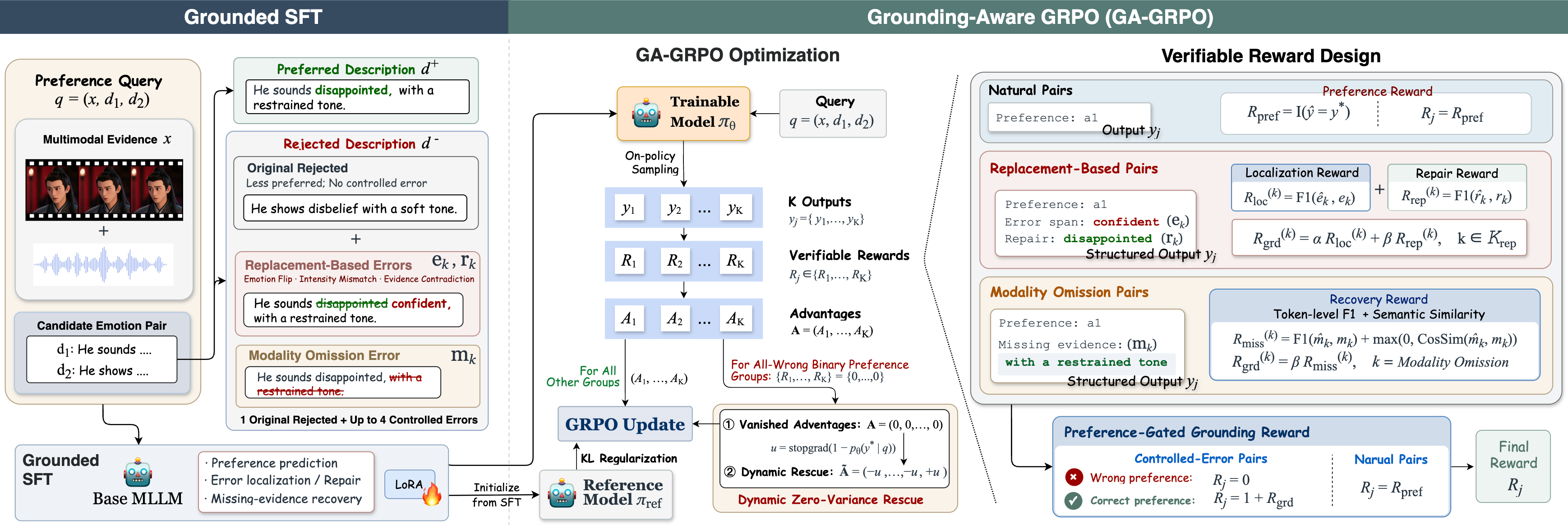}
\caption{Overview of the proposed EmoGavel framework for multimodal emotion preference learning. Grounded SFT learns preference prediction and error grounding from natural and controlled-error pairs, and it initializes the trainable and reference policies. GA-GRPO samples multiple outputs, computes verifiable rewards and group-relative advantages, applying zero-variance rescue to all-wrong binary preference groups and updating the policy with KL regularization. The reward function uses preference-only rewards for natural pairs and
preference-gated grounding rewards for controlled-error pairs.}
\label{fig:overview}
\end{figure*}

\section{Method}
\label{sec:method}
\subsection{Overall Architecture}
As shown in Fig.~\ref{fig:overview}, EmoGavel is a two-stage framework that extends multimodal emotion preference learning with explicit error grounding. In addition to predicting which candidate description is preferred, the model is trained to identify and correct erroneous content when verifiable error information is available. Grounded SFT first learns this joint preference--grounding task from supervised targets, and Grounding-Aware GRPO (GA-GRPO) subsequently optimizes the corresponding verifiable outcomes.

\subsection{Task Formulation and Grounding Supervision}

Given a multimodal clip $x$ and two candidate descriptions $(d_1,d_2)$, conventional emotion preference prediction asks a judge to determine which description better reflects the target person's emotional state. We denote the input as $q=(x,d_1,d_2)$ and the ground-truth preference
label as $y^*\in\{a_1,a_2\}$, where $a_i$ selects $d_i$~\cite{lian2026emoprefer}.
We distinguish two types of preference pairs. A natural pair $(d^+,d^-)$ contains a human-preferred description $d^+$ and a rejected description $d^-$, where $d^-$ is less preferred
but not necessarily erroneous. In contrast, EAPO constructs controlled-error pairs $(d^+,d_k^-)$ by introducing one of four predefined errors into the preferred description: \emph{Emotion Flip}, \emph{Intensity Mismatch}, \emph{Evidence Contradiction}, and \emph{Modality Omission}~\cite{huang2026eapo}. These controlled negatives were originally used to enrich binary preference training, but their construction process also preserves how the preferred description is transformed into the wrong one, providing supervision beyond the preference label.

We therefore augment the conventional preference task with an
error-grounding task on controlled pairs. Let
$\mathcal K_{\mathrm{rep}}$ denote the three replacement-based errors:
\emph{Emotion Flip}, \emph{Intensity Mismatch}, and
\emph{Evidence Contradiction}, and define
$\mathcal K=\mathcal K_{\mathrm{rep}}\cup
\{\text{Modality Omission}\}$ as the four controlled-error types.
For $k\in\mathcal K_{\mathrm{rep}}$, the modified span defines the
localization target $e_k$, while the corresponding original span defines
the repair target $r_k$. 
For $k=\text{Modality Omission}$, the removed content defines the
recovery target $m_k$. As illustrated in Fig.~\ref{fig:overview}, replacing the ground-truth ``disappointed''
with the wrong``confident'' gives $e_k=\text{``confident''}$ and
$r_k=\text{``disappointed''}$, while removing ``with a restrained tone''
gives the recovery target $m_k=\text{``with a restrained tone''}$. These targets are directly derived from the controlled-error
construction process and require no additional manual annotation.
Natural pairs retain only preference supervision, whereas controlled
pairs provide both preference and grounding supervision.

\subsection{Grounded Supervised Fine-Tuning (Grounded SFT)}

Conventional preference SFT learns only the final $a1/a2$ decision. Grounded SFT additionally learns structured error grounding from the
targets defined above. Natural and controlled binary pairs retain the standard $a1/a2$ preference target, while controlled errors additionally provide structured grounding supervision.
For the replacement-based error $k\in\mathcal K_{\mathrm{rep}}$, the target structured model output is:
\begin{center}
\begin{minipage}{0.90\columnwidth}\small\ttfamily
Preference: a1\\
Error span: $e_k$ (corrupted span)\\
Repair: $r_k$ (original span)
\end{minipage}
\end{center}
For modality omission error $k=\text{Modality Omission}$, the target model output becomes:
\begin{center}
\begin{minipage}{0.90\columnwidth}\small\ttfamily
Preference: a1\\
Missing evidence: $m_k$ (removed content)
\end{minipage}
\end{center}
Thus, Grounded SFT jointly learns preference prediction, error
localization and repair, or missing-evidence recovery. Given an input
$q$ and target output $t=(t_1,\ldots,t_L)$, we minimize the standard
autoregressive cross-entropy:
\begin{equation}
\Lcal_{\mathrm{Grounded,SFT}}
=
-\frac{1}{L}\sum_{\ell=1}^{L}
\log p_\theta(t_\ell \mid q,t_{<\ell}),
\label{eq:gsft_nll}
\end{equation}
where $L$ is the number of target tokens, $t_\ell$ denotes the $\ell$-th
ground-truth token, and $t_{<\ell}$ denotes its preceding target tokens.
The prompt and multimodal input tokens are masked from the loss. Binary and structured examples therefore share the same training objective and differ only in their target format.

For grounded SFT, we adopt low-rank adaptation (LoRA)~\cite{hu2022lora}, and the resulting adapter is used to initialize both the trainable policy $\pi_\theta$ and the frozen reference policy $\pi_{\mathrm{ref}}$ in GA-GRPO. Through Grounded SFT, the model acquires preference prediction and error-grounding capabilities, while the subsequent
RL stage directly optimizes task-level outcomes.

\subsection{Grounding-Aware GRPO (GA-GRPO)}

\subsubsection{Verifiable Preference and Grounding Rewards}
Our controlled-error construction provides explicit, verifiable
targets for both preference and error grounding, allowing rewards
to be computed directly without an additional learned reward model.

Let $\hat y$ denote the parsed preference prediction and
$y^*$ the ground-truth preference label. The preference reward is defined as
\begin{equation}
R_{\mathrm{pref}}(\hat y,y^*)=
\begin{cases}
1, & \hat y=y^*,\\
0, & \text{otherwise}.
\end{cases}
\label{eq:pref_reward}
\end{equation}

For $k\in\mathcal K_{\mathrm{rep}}$, the controlled-error construction
provides the erroneous span $e_k$ and its corresponding original span
$r_k$. We measure localization and repair rewards using token-level F1 score:
\begin{equation}
R_{\mathrm{loc}}^{(k)}
=
\mathrm{F1}(\hat e_k,e_k),
\qquad
R_{\mathrm{rep}}^{(k)}
=
\mathrm{F1}(\hat r_k,r_k),
\label{eq:loc_rep_reward}
\end{equation}
where $\hat e_k$ and $\hat r_k$ denote the predicted error span and
repair, respectively.
For $k=\text{Modality Omission}$, the construction instead provides the
removed evidence $m_k$. Since the same missing emotional evidence may allow multiple
semantically equivalent expressions, we define the recovery reward as the sum of token-level F1 and frozen-encoder semantic similarity ~\cite{reimers2019sentencebert}:
\begin{equation}
R_{\mathrm{miss}}^{(k)}
=
\mathrm{F1}(\hat m_k,m_k)
+
\max\!\left(0,\mathrm{CosSim}(\hat m_k,m_k)\right)
,
\end{equation}
where $\mathrm{CosSim}(\cdot,\cdot)$ denotes cosine similarity computed
using the frozen sentence encoder.
The grounding reward is then defined according to the controlled-error
type:
\begin{equation}
R_{\mathrm{grd}}^{(k)}
=
\begin{cases}
\alpha R_{\mathrm{loc}}^{(k)}
+\beta R_{\mathrm{rep}}^{(k)},
& k\in\mathcal K_{\mathrm{rep}},\\[2mm]
\beta R_{\mathrm{miss}}^{(k)},
& k=\text{Modality Omission}.
\end{cases}
\label{eq:grounding_reward}
\end{equation}

Finally, grounding reward is gated by preference correctness. For a controlled-error example, the total reward is
\begin{equation}
R=
\begin{cases}
0, & R_{\mathrm{pref}}=0,\\
1+R_{\mathrm{grd}}^{(k)}, & R_{\mathrm{pref}}=1.
\end{cases}
\label{eq:gated_reward}
\end{equation}
For the $j$-th sampled output, we denote the resulting total reward
by $R_j$. Natural preference examples contain no verifiable grounding target
and therefore use $R_j=R_{\mathrm{pref}}$. This design imposes a preference-first reward hierarchy: grounding rewards are activated only when the preference prediction is correct.

\subsubsection{Dynamic Zero-Variance Rescue}

For each prompt $q$, GRPO samples $K$ outputs
$\{y_1,\ldots,y_K\}$ and computes group-relative advantages. Rewards are normalized within the group to obtain
the scalar advantage of each output~\cite{shao2024deepseekmath}:
\begin{equation}
A_j=
\frac{R_j-\operatorname{mean}(R_1,\ldots,R_K)}
{\operatorname{std}(R_1,\ldots,R_K)+\epsilon},
\qquad j=1,\ldots,K,
\label{eq:grpo_advantage}
\end{equation}
where $R_j$ is the reward of the $j$-th output and $\epsilon>0$ is a small constant for numerical stability.
The corresponding group advantage vector is
$\mathbf{A}=(A_1,\ldots,A_K)$. 

When all $K$ outputs are incorrect, their rewards are all zero,
yielding $\mathbf{A}=(0,\ldots,0)$ and thus no corrective preference
signal. 
We therefore introduce a dynamic rescue for such all-wrong binary
groups. Let $y^*$ denote the ground-truth preference label and define
\begin{equation}
u=
\operatorname{stopgrad}
\left(1-p_\theta(y^*\mid q)\right),
\label{eq:difficulty}
\end{equation}
where $p_\theta(y^*\mid q)$ is the normalized probability of the
ground-truth label over $\{\texttt{a1},\texttt{a2}\}$, computed from
their sequence log-likelihoods. The adaptive score $u$ increases when
the model assigns lower confidence to the correct preference. For an all-wrong binary preference group, we append the ground-truth
output $y^*$ to the $K$ incorrect outputs, forming
\begin{equation}
\mathcal{Y}_{\mathrm{rescue}}
=
\{y_1^-,\ldots,y_K^-,y^*\}.
\label{eq:rescued_group}
\end{equation}
The rescued advantage vector is defined as
\begin{equation}
\widetilde{\mathbf A}
=
(\underbrace{-u,\ldots,-u}_{K\ \text{incorrect outputs}},+u),
\label{eq:zv_advantage}
\end{equation}
where the final positive advantage $+u$ corresponds to the auxiliary ground-truth output $y^*$. Thus, incorrect preference outputs are
suppressed while the ground-truth label is reinforced.
The rescue is applied only to all-wrong binary preference groups;
all other groups retain the original GRPO advantage vector
$\mathbf A=(A_1,\ldots,A_K)$. We denote the resulting objective as
$\mathcal{L}_{\mathrm{GA\text{-}GRPO}}$, which follows the GRPO update for standard groups and applies the rescue-specific augmented update to all-wrong binary groups, with KL regularization to the frozen Grounded SFT reference policy $\pi_{\mathrm{ref}}$.
\section{Experiments}
\label{sec:experiments}

\subsection{Experimental Setup}

EmoPrefer provides pairwise preference data for multimodal emotion
descriptions, with EmoPrefer-Data-V2 used for development and
EmoPrefer-Data serving as the held-out benchmark~\cite{lian2026emoprefer}.
Following EAPO~\cite{huang2026eapo}, both sets are augmented with up to
four controlled-error descriptions derived from each preferred
description. Each video-level group therefore contains one preferred
description, one natural rejected description, and up to four
controlled-error variants.

We split the EAPO-extended EmoPrefer-Data-V2 into 90\% training and
10\% validation sets by video ID to prevent source-level leakage.
Final evaluation uses EmoPrefer-Data for natural preference and its
EAPO extension for controlled-error preference, swap consistency, and
error grounding. We report weighted F1 (WAF) on natural pairs, mean WAF over the four
controlled-error types, and candidate-order consistency. Parse is the percentage
of successfully parsed structured outputs, with malformed outputs
counted as failures. Grounding Macro averages the three replacement-error
scores (mean localization and repair F1) and the Modality Omission
recovery score, which reports the average of token-level F1 and frozen-encoder cosine similarity, scaled to percentage.

We use Qwen2.5-Omni-7B as the base multimodal LLM. Following the S2
setting of EmoPrefer~\cite{lian2026emoprefer}, the model receives an
auxiliary multimodal description together with video, audio,
and the two candidates. Grounded SFT is initialized from a
natural-preference checkpoint and trained for two epochs using a 50/50
mixture of binary preference and structured grounding examples. The
structured examples are balanced across the four controlled-error types,
with candidate swapping applied at probability 0.5. We use LoRA with
rank 8, scaling factor 16, and dropout 0.05, with learning rates of
$5\times10^{-6}$ and $2.5\times10^{-6}$ for the first and second
epochs, respectively. GA-GRPO samples $K=4$ outputs with temperature
0.9 and top-$p$ 0.95, and is trained for 500 steps with a learning rate
of $1\times10^{-7}$. We set $\epsilon=10^{-4}$ for reward
normalization, $\epsilon_{\mathrm{clip}}=0.2$ for GRPO clipping,
$\alpha=\beta=0.25$, and the KL coefficient to 0.12. The final checkpoint is selected by validation 4-Error Avg WAF. Our
Qwen2.5-Omni training experiments are conducted on eight NVIDIA RTX
A6000 GPUs.
\begin{table*}[t]
\caption{Preference performance (\%) on EmoPrefer-Data and its EAPO
extension. The column “Natural" shows the WAF on natural pairs, while
“4-Error Avg" and “Swap Cons" report the mean WAF and candidate-order
consistency over the four controlled-error types, respectively.
 Each Judge ID identifies a model with its inference setting (S2) and optimization strategy. SFT uses standard preference supervision, whereas Error-Aug SFT denotes preference-only SFT using both natural and
EAPO controlled-error pairs.~\cite{huang2026eapo}. Best and second-best results are shown in
\textbf{bold} and \underline{underline}, respectively.}
\label{tab:main_results}
\vspace{4pt}
\centering
\tiny
\setlength{\tabcolsep}{2.5pt}
\renewcommand{\arraystretch}{0.94}
\resizebox{\textwidth}{!}{%
\begin{tabular}{cllrrrrrrr}
\toprule
\multicolumn{3}{c}{\textbf{Judge Configuration}} & \multicolumn{1}{c}{} & \multicolumn{4}{c}{\textbf{4 Controlled-Error Subsets}} & \multicolumn{1}{c}{} & \multicolumn{1}{c}{} \\
\cmidrule(lr){1-3}\cmidrule(lr){5-8}
\textbf{ID} & \textbf{Model} & \textbf{Setting / Method} & \cellcolor{metricgray}\textbf{Natural} & \textit{Emotion Flip} & \textit{Intensity Mismatch} & \textit{Evidence Contradiction} & \textit{Modality Omission} & \cellcolor{metricgray}\textbf{4-Error Avg} & \textbf{Swap Cons} \\
\midrule
1 & Qwen3-Omni-30B-Thinking~\cite{xu2025qwen3omni} & S2 Zero-shot & \cellcolor{metricgray}73.36 & 92.27 & 54.51 & 71.20 & 65.95 & \cellcolor{metricgray}70.98 & 62.29 \\
2 & Qwen3-Omni-30B-Instruct~\cite{xu2025qwen3omni} & S2 Zero-shot & \cellcolor{metricgray}{73.43} & 87.06 & 52.29 & 63.49 & 77.75 & \cellcolor{metricgray}{70.15} & 54.13 \\
3 & Qwen3-Omni-30B-Instruct & S2 Error-Aug SFT & \cellcolor{metricgray}{76.56} & 92.74 & 61.64 & 71.48 & 76.79 & \cellcolor{metricgray}{75.66} & 65.47 \\
4 & MiniCPM~\cite{openbmb2025minicpmo} + Qwen2.5 + Qwen3 & EAPO~\cite{huang2026eapo} & \cellcolor{metricgray}\textbf{80.31} & \textbf{96.21} & \underline{77.01} & \underline{75.46} & \textbf{92.72} & \cellcolor{metricgray}\underline{85.35} & \underline{76.38} \\
\midrule
5 & Qwen2.5-Omni-7B-Instruct~\cite{xu2025qwen25omni} & S2 Zero-shot & \cellcolor[HTML]{ececec}{68.65} & 83.80 & 49.30 & 51.82 & 47.59 & \cellcolor[HTML]{ececec}{58.01} & 34.96 \\
6 & Qwen2.5-Omni-7B-Instruct & S2 SFT & \cellcolor[HTML]{ececec}{77.08} & 86.44 & 51.35 & 62.57 & 80.48 & \cellcolor[HTML]{ececec}{70.21} & 53.18 \\

7 & Qwen2.5-Omni-7B-Instruct & S2 Error-Aug SFT & \cellcolor[HTML]{ececec}{77.15} & 94.32 & 67.13 & 68.58 & 78.49 & \cellcolor[HTML]{ececec}{77.13} & 64.41 \\
8 & Qwen2.5-Omni-7B-Instruct & S2 Grounded SFT
& \cellcolor{metricgray}78.13
& 92.88
& 62.62
& 68.78
& 92.46
& \cellcolor{metricgray}79.18
& 68.11 \\
\midrule
9 & Qwen2.5-Omni-7B-Instruct & \textbf{\method} & \cellcolor{metricgray}\underline{79.75} & \underline{95.90} & \textbf{83.73} & \textbf{80.70} & \underline{92.47}
& \cellcolor{metricgray}\textbf{88.20} & \textbf{81.25} \\

\bottomrule
\end{tabular}%
}
\par\vspace{-10pt}
\end{table*}
\begin{table*}[t]
\caption{Error-grounding performance (\%) across training strategies.
“Loc." and “Rep." denote token-level F1 scores for localization and repair, respectively. “Evid. Recovery" evaluates missing-evidence recovery for Modality Omission. “Parse" denotes the rate of correctly formatted
outputs. “Natural" and “4-Error Avg" report preference WAF on natural pairs
and the mean over four controlled-error types, respectively. “Grounding"
denotes the macro grounding score. Best and second-best results are shown
in \textbf{bold} and \underline{underline}, respectively.}
\label{tab:grounding_detail}
\vspace{4pt}
\centering
\scriptsize
\setlength{\tabcolsep}{2.7pt}
\renewcommand{\arraystretch}{1.08}
\resizebox{\textwidth}{!}{%
\begin{tabular}{l
                r
                *{6}{>{\centering\arraybackslash}p{30pt}}
                c
                rrr}
\toprule
\multicolumn{2}{c}{Qwen2.5-Omni-7B-Instruct}
& \multicolumn{2}{c}{\textit{Emotion Flip}}
& \multicolumn{2}{c}{\textit{Intensity Mismatch}}
& \multicolumn{2}{c}{\textit{Evidence Contradiction}}
& \multicolumn{1}{c}{\textit{Modality Omission}}
& \multicolumn{3}{c}{\textbf{Overall}} \\
\cmidrule(lr){3-4}
\cmidrule(lr){5-6}
\cmidrule(lr){7-8}
\cmidrule(lr){9-9}
\cmidrule(lr){10-12}

\textbf{Variant}
& \textbf{Parse}
& \textbf{Loc.} & \textbf{Rep.}
& \textbf{Loc.} & \textbf{Rep.}
& \textbf{Loc.} & \textbf{Rep.}
& \textbf{Evid. Recovery}
& \cellcolor{metricgray}\textbf{Natural}
& \cellcolor{metricgray}\textbf{4-Error Avg}
& \cellcolor{metricgray}\textbf{Grounding} \\
\midrule

Zero-shot
& 87.00
& 27.09 & 25.45
& 48.33 & 39.66
& 49.24 & 46.30
& 18.70
& \cellcolor{metricgray}68.65
& \cellcolor{metricgray}58.01
& \cellcolor{metricgray}34.18 \\

SFT
& 90.06
& 31.19 & 30.20
& 60.42 & 53.31
& 52.76 & 51.68
& 26.78
& \cellcolor{metricgray}77.08
& \cellcolor{metricgray}70.21
& \cellcolor{metricgray}41.64 \\

Error-Aug SFT
& \underline{91.51}
& 35.12 & 34.50
& 62.66 & 57.56
& 56.81 & 55.04
& 29.61
& \cellcolor{metricgray}77.15
& \cellcolor{metricgray}77.13
& \cellcolor{metricgray}45.11 \\

\midrule

Grounded SFT
& 89.73
& \underline{78.43} & \underline{78.21}
& \underline{78.29} & \underline{75.42}
& \underline{72.29} & \underline{72.24}
& \textbf{63.91}
& \cellcolor{metricgray}\underline{78.13}
& \cellcolor{metricgray}\underline{79.18}
& \cellcolor{metricgray}\underline{72.84} \\

\textbf{\method}
& \textbf{98.42}
& \textbf{92.39} & \textbf{92.37}
& \textbf{80.14} & \textbf{78.42}
& \textbf{76.30} & \textbf{75.97}
& \underline{62.51}
& \cellcolor{metricgray}\textbf{79.75}
& \cellcolor{metricgray}\textbf{88.20}
& \cellcolor{metricgray}\textbf{77.27} \\

\bottomrule
\end{tabular}
}
\par\vspace{-10pt}
\end{table*}

\begin{table}[t]
\caption{Ablation results for GA-GRPO. Best and second-best results are
shown in \textbf{bold} and \underline{underline}, respectively.}
\label{tab:ablation}
\vspace{4pt}
\centering
\scriptsize
\setlength{\tabcolsep}{5.0pt}
\renewcommand{\arraystretch}{1.08}
\begin{tabular}{lrrr}
\toprule
\textbf{Variant}
& \textbf{Natural}
& \textbf{4-Error Avg}
& \textbf{Grounding} \\
\midrule

Grounded SFT
& 78.13
& 79.18
& 72.84 \\
\midrule

Vanilla GRPO
& 78.53
& 81.13
& 72.73 \\

\text{ } + Grounding Reward
& 78.38
& 81.53
& 76.32 \\

\text{ }  + Grounding Reward + Dynamic Rescue
& \underline{79.03}
& 82.11
& \textbf{78.12} \\

\text{ } + Grounding Reward + Preference Gate
& 78.88
& \underline{85.62}
& 76.99 \\

\rowcolor{metricgray}
\textbf{GA-GRPO (Full)}
& \textbf{79.75}
& \textbf{88.20}
& \underline{77.27} \\
\bottomrule
\end{tabular}
\par\vspace{-10pt}
\end{table}
\vspace{-12pt}
\subsection{Analysis of Preference Robustness}

Table~\ref{tab:main_results} shows that stronger performance on natural
pairs does not necessarily imply robustness to controlled errors.
Standard SFT improves both natural and controlled-error preference
prediction, while Error-Aug SFT and Grounded SFT provide further gains
predominantly on the controlled-error subsets.
EmoGavel reaches 79.75\% WAF on the natural set, the best result among all
single-model judges, while achieving the best 4-Error Avg of 88.20\%
and Swap Cons of 81.25\%. EAPO remains slightly higher on the natural
set through fusion of three adapted judges~\cite{huang2026eapo},
whereas EmoGavel attains stronger controlled-error robustness and order
consistency with a single Qwen2.5-Omni-7B judge.

\subsection{Analysis of Error Grounding}

Table~\ref{tab:grounding_detail} exposes the limitation of
preference-only training: selecting the better description does not
necessarily mean that the model can identify where the alternative is
wrong or how it should be corrected. Although SFT and Error-Aug SFT
already improve preference prediction and output-format compliance,
their localization and repair performance remains limited. Explicit
grounding supervision substantially closes this gap, and GA-GRPO further
raises the overall Grounding score to 77.27\%. However, its Evid. Recovery
on Modality Omission drops below Grounded SFT, suggesting that the current
GA-GRPO objective is better aligned with replacement-based localization
and repair than with omission recovery. Improving this balance remains an
important direction for the reward design.

\subsection{Ablation Studies on GA-GRPO}

Table~\ref{tab:ablation} shows that vanilla GRPO provides only modest
improvements over Grounded SFT, indicating that preference reward alone
is insufficient to substantially enhance error grounding. Adding the
grounding reward raises Grounding from 72.73\% to 76.32\% with little
change in preference performance.

Dynamic rescue further improves Grounding to 78.12\%, while the
preference gate yields a larger gain in controlled-error robustness,
increasing 4-Error Avg from 81.53\% to 85.62\%. Combining both
components achieves the strongest overall preference performance,
reaching 79.75\% Natural and 88.20\% 4-Error Avg while maintaining
77.27\% Grounding. These results suggest that grounding reward,
dynamic rescue, and preference gating provide complementary benefits.

\FloatBarrier
\section{Conclusion}
\label{sec:conclusion}

We propose EmoGavel, a grounding-aware framework for robust multimodal
emotion preference learning that goes beyond identifying human-preferred
emotion descriptions by enabling verifiable error grounding. EmoGavel employs Grounded SFT to learn preference prediction together with error localization, repair, and missing-evidence recovery, and then introduces GA-GRPO to jointly optimize preference and grounding through preference-gated rewards and zero-variance rescue. Experiments
show that a single Qwen2.5-Omni-7B judge achieves 79.75\% WAF on natural preference pairs, 88.20\% average WAF across 4 controlled-error types, and 72.27\% Grounding, demonstrating a strong
balance between preference prediction, controlled-error robustness, and error-grounding capability. These results show that controlled-error supervision supports
both preference prediction and verifiable error grounding. Future work will extend the current controlled-error setting
to more naturally occurring and less explicitly aligned errors, and
further examine grounding under different multimodal evidence conditions.
\clearpage
\bibliographystyle{IEEEbib}
\bibliography{refs}

\end{document}